\documentclass[fleqn,usenatbib]{mnras}

\usepackage{newtxtext,newtxmath}

\usepackage[T1]{fontenc}

\DeclareRobustCommand{\VAN}[3]{#2}
\let\VANthebibliography\thebibliography
\def\thebibliography{\DeclareRobustCommand{\VAN}[3]{##3}\VANthebibliography}

\usepackage{amsmath}
\usepackage{bm}
\usepackage{tikz}
\usetikzlibrary{shapes.geometric, arrows, pgfplots.groupplots}
\tikzstyle{process} = [rectangle, minimum width=3cm, minimum height=1cm, text centered, draw=black]
\tikzstyle{start} = [rectangle, rounded corners, minimum width=3cm, minimum height=1cm,text centered, draw=black]
\tikzstyle{arrow} = [thick,->,>=stealth]

\usepackage{pgfplots}
\pgfplotsset{
    compat=1.16,
    legend image with text/.style={
        legend image code/.code={%
            \node[anchor=center] at (0.3cm,0cm) {#1};
        }
    },
}

\definecolor{cbdark2-1}{RGB}{27,158,119}
\definecolor{cbdark2-2}{RGB}{217,95,2}
\definecolor{cbdark2-3}{RGB}{117,112,179}
\definecolor{cbdark2-4}{RGB}{231,41,138}
\definecolor{cbdark2-5}{RGB}{102,166,30}
\definecolor{cbdark2-6}{RGB}{230,171,2}
\definecolor{cbdark2-7}{RGB}{166,118,29}
			
\pgfplotscreateplotcyclelist{colour list corr}{%
cbdark2-1, only marks, mark=*\\%
cbdark2-1, dashed\\%
cbdark2-2,only marks, mark=square*\\%
cbdark2-2,dashed\\%
cbdark2-3,only marks, mark=otimes\\%
cbdark2-3,dashed\\%
cbdark2-4,only marks, mark=triangle*\\%
cbdark2-4,dashed\\%
cbdark2-5,only marks, mark=diamond*\\%
cbdark2-5,dashed\\%
cbdark2-6,only marks, mark=star\\%
cbdark2-6,dashed\\%
cbdark2-7,only marks, mark=square*\\%
cbdark2-7,dashed\\%
}

\pgfplotscreateplotcyclelist{colour list times}{%
cbdark2-1, dashed, mark=*\\%
cbdark2-1, mark=triangle\\%
cbdark2-2, dashed, mark=*\\%
cbdark2-2, mark=triangle\\%
}

\pgfplotscreateplotcyclelist{colour list times all}{%
cbdark2-1, mark=*\\%
cbdark2-2, mark=square*\\%
cbdark2-3, mark=otimes\\%
cbdark2-4, mark=triangle*\\%
cbdark2-7, mark=diamond*\\%
}

\pgfplotscreateplotcyclelist{colour list params}{%
cbdark2-1, dashed, mark=*\\%
cbdark2-2, mark=triangle\\%
}

\newcommand{\routine}[1]{\textsc{\lowercase{#1}}}
\newcommand{\parameter}[1]{\textsc{\lowercase{#1}}}
\newcommand{\program}[1]{\textsc{\lowercase{#1}}}
\newcommand{\NBODY}{\program{NBODY6}}
\newcommand{\NBPPPT}{\program{NBODY6+P3T}}

\title[Accelerating NBODY6 with a GPU-Enabled P3T Scheme]{Accelerating NBODY6 with a GPU-Enabled Particle-Particle Particle-Tree Scheme}

\author[A. D. Arnold et al.]{
Anthony D. Arnold,$^{1}$\thanks{E-mail: anthony.arnold@uqconnect.edu.au (ADA)}
Holger Baumgardt$^{1}$
and Long Wang$^{2,3}$
\\
$^{1}$School of Mathematics and Physics, The University of Queensland, St Lucia QLD 4072, Australia\\
$^{2}$Department of Astronomy, School of Science, The University of Tokyo, 7-3-1 Hongo, Bunkyo-ku, Tokyo 113-0033, Japan\\
$^{3}$RIKEN Center for Computational Science, Minatojima-minami-mach, Chuo-ku, Kobe, Hyogo 650-0047, Japan
}

\date{Accepted XXX. Received YYY; in original form ZZZ}

\pubyear{2021}

\begin{document}
\label{firstpage}
\pagerange{\pageref{firstpage}--\pageref{lastpage}}
\maketitle

\begin{abstract}
We describe a modified version of the \NBODY~code for simulating star clusters which greatly improves computational efficiency while sacrificing little in the way of accuracy. The distant force calculator is replaced by a GPU-enabled Barnes-Hut code, and integration is done with a standard leap frog scheme. Short-range forces continue to use the CPU-based fourth-order Hermite predictor-corrector scheme of \NBODY. Our code outperforms \NBODY~for systems with more than $3 \times10^5$ particles and runs more than a factor $2$ faster for systems of $10^6$ particles with similar energy conservation. Our code should be useful for simulating realistic dense stellar clusters, such as globular clusters or galactic nuclei.
\end{abstract}




\section{Introduction}

The \NBODY~code~\citep{Aarseth99} is the state of the art code for performing direct N-body simulations to study the evolution of dense stellar systems. It is the yardstick against which other N-body codes are measured in terms of computational efficiency and accuracy~\citep{iwasawa2015gpuenabled,2016MNRAS.463.2109R}. Moreover, \NBODY~remains flexible in its support for arbitrary initial conditions compared to Monte-Carlo methods~\citep{2013MNRAS.431.2184G}. The code splits the gravitational force acting on a particle into near and distant forces, allowing the fast-changing near forces to be treated separately from the slow-changing distant ones~\citep{1973JCoPh..12..389A,1992PASJ...44..141M}. Recent improvements to \NBODY~have included routines for performing distant force calculations on general-purpose graphics processing units (GPUs)~\citep{2012MNRAS.424..545N}, which greatly improves the overall run time. Since the force calculation algorithm used in \NBODY~has a calculation cost of $O(N^2)$, simulations of large clusters, exceeding $10^6$ stars, are still very expensive~\citep{iwasawa2015gpuenabled,Bonsai,aarseth_2003,2015MNRAS.450.4070W}.

An early attempt at an $O(N \log N)$ integration scheme for N-body simulations using a tree code was described by~\citet{1993ApJ...414..200M}. More recently,~\citet{Oshino_2011} described the Particle-Particle Particle-Tree (P3T) scheme which uses a tree code for distant force calculations and a standard leap-frog scheme for integrating these distant forces, while retaining the Hermite scheme for neighbour forces. A working implementation of this scheme with a GPU-enabled tree code was shown in~\citet{iwasawa2015gpuenabled}. Over short simulation times, the P3T scheme performs faster than \NBODY, however as the system evolves past core collapse, the P3T scheme becomes slow due to the short time steps required by the Hermite integrator. \NBODY~is able to cope with these short time steps due to the KS~\citep{KSReg} and chain regularisation code for close encounters~\citep{MikkolaAarseth89,2003IAUS..208..295A}. Additionally, \NBODY~includes routines for modelling stellar evolution and post-Newtonian force corrections, which allow for more realistic simulations.

In this paper, we present a version of \NBODY, called \NBPPPT, which incorporates the P3T scheme while retaining \NBODY's routines for KS and chain regularisation. Section~\ref{construction} describes the construction of the improved \NBODY~code. Section~\ref{results} presents the results of different runs compared to the original \NBODY. Finally in section~\ref{conclusions} we present conclusions and discuss future applications of the new code.

\section{Implementation}
\label{construction}
\subsection{Formulation}

In this section we describe the modifications made to \NBODY~in order to incorporate the P3T scheme. We adopt the definition of P3T from~\citet{iwasawa2015gpuenabled} with the following deviations. The Plummer softening is omitted from the scheme as the singularity due to the $1/r$ potential is handled by special treatment of close encounters through KS regularisation~\citep{MIKKOLA1998309}. 

A radius known as the neighbour sphere around each particle separates the force acting on the particle into irregular and regular forces. The irregular forces are those fast-changing forces from nearby particles inside the neighbour sphere. The regular forces are from the distant particles outside of the neighbour sphere, which change more slowly. In \NBPPPT, the irregular and regular accelerations of particle $i$, $\bm{F}_{I,i}$ and $\bm{F}_{R,i}$ respectively, due to particle $j$ are calculated as follows:
\begin{align}
    \label{eq:fi}
    \bm{F}_{I,i} &= \sum_{j \ne i}^{N} m_{j} \frac{\bm{r}_{ij}}{|\bm{r}_{ij}|^{3}} K_{ij},\\
    \label{eq:fidot}
    \dot{\bm{F}}_{I,i} &= \sum_{j \ne i}^{N} m_{j}\frac{\bm{\dot r}_{ij}}{|\bm{r}_{ij}|^{3}} K_{ij} - 3 \frac{(\bm{r}_{ij} \cdot \bm{\dot r}_{ij}) \bm{r}_{ij}}{|\bm{r}_{ij}|^{5}}  + m_{j} \frac{\bm{r}_{ij}}{|\bm{r}_{ij}|^{3}} K'_{ij},\\
    \bm{F}_{R,i} &= \sum_{j \ne i}\frac{m_j}{|\bm{r}_{ij}|^{3}}\bm{r}_{ij} - \bm{F}_{I,i},\\
    \bm{r}_{ij} &= \bm{x}_i - \bm{x}_j,\\
    \bm{\dot r}_{ij} &= \bm{\dot x}_i -  \bm{\dot x}_j
\end{align}%
where $m_i$, $\bm{x}_i$, $\bm{\dot x}_i$ are the mass, position, and velocity of particle $i$, respectively and $\bm{r}_{ij}$ is the Euclidean distance between particles $i$ and $j$. Here, $K$ and $K'$ are cutoff functions to ensure a smooth transition of forces between $\bm{F}_{I}$ and $\bm{F}_{R}$ as particles move in and out of neighbour spheres. We adopt the formulae from~\citet{iwasawa2015gpuenabled,Duncan98amultiple}, we have:
\begin{align}
    \label{eq:duncan}
    K_{ij} &= 1 - \begin{cases}
               0 &\text{if }x<0,\\
               -20x^7 + 70x^6 - 84x^5 + 35x^4 &\text{if } 0 \leq x < 1,\\
               1 &\text{if }1 \leq x,\\
            \end{cases}\\
    \label{eq:duncan2}
    K'_{ij} &= \begin{cases}
               (-140x^6 + 420x^5 - 420x^4 + 140x^3)\frac{(\bm{r}_{ij} \cdot \bm{\dot r}_{ij})}{|\bm{r}_{ij}| ({r_{\text{cut}}}-r_{\text{in}})} &\text{if }0 \leq x < 1,\\
               0 &\text{otherwise,}
            \end{cases}
\end{align}%
where%
\begin{align}
    x &= \frac{y - \gamma}{1 - \gamma},\\
    y &= \frac{|\bm{r}_{ij}|}{r_{\text{cut}}},\\
    \gamma &= \frac{r_{\text{in}}}{r_{\text{cut}}},
\end{align}%
and $r_{\text{cut}}$ and $r_{\text{in}}$ denote the outer and inner cutoff radii for neighbour forces, respectively. If $|\bm{r}_{ij}| < r_{\text{in}}$ then the force from particle $j$ contributes entirely to $\bm{F}_{I,i}$, and likewise if $|\bm{r}_{ij}| > r_{\text{cut}}$ the force counts wholly as part of $\bm{F}_{R,i}$. As in~\citet{iwasawa2015gpuenabled}, we use $\gamma = 0.1$ for all calculations.

In \NBPPPT, we used the standard criterion from~\citet{1992PASJ...44..141M}, which determines the individual time step $\Delta t_i$ for each particle in the Hermite scheme.
\begin{equation}
    \label{eq:dt}
    \Delta t_{i} = \min \left( \sqrt{\eta
            \frac
            {\left|\bm a_i^{(0)}\right| \left|\bm a_i^{(2)}\right| + \left|\bm a_i^{(1)}\right|^2 }
            {\left|\bm a_i^{(1)}\right| \left|\bm a_i^{(3)}\right| + \left|\bm a_i^{(2)}\right|^2}
        }, \Delta t_{\text{max}} \right).
\end{equation}%
Here, $\bm a_i^{(n)}$ is the $N\text{th}$ time derivative of the acceleration of particle $i$ due to $\bm{F}_{I,i}$, with the exception of $\bm a_i^{(0)}$ which is due to $\bm{F}_{I,i} + \bm{F}_{R,i}$~\citep{1992PASJ...44..141M}. We used the suggestion from~\citet{1985mts..conf..377A} and set the accuracy parameter $\eta = 0.02$ for all calculations.

The standard P3T implementation includes an initial time step criterion to account for the missing higher order acceleration derivatives. \NBODY~instead has routines to calculate these derivatives and as such the fourth order standard criterion is used to calculate the initial time step in \NBPPPT.

\subsection{Modified Integration}

In \NBPPPT~we replaced the regular force Hermite scheme from \NBODY~with a standard leapfrog integrator. The velocity kick from the regular force is given by%
\begin{equation}
    \label{eq:kick}
    \bm{\dot x}_i = \bm{\dot x}_i + \frac{\Delta t_{\text{reg}}}{2} \bm{F}_{R,i}.
\end{equation}%
The regular time step $\Delta t_{\text{reg}}$ is a constant shared by all particles and is defined as a multiple of the maximum irregular Hermite step $\Delta t_{\text{max}}$. In all of our simulations we use $\Delta t_{\text{reg}} = 4 \Delta t_{\text{max}}$. The integration algorithm used in \NBPPPT, illustrated in Fig.~\ref{fig:intgrt}, is as follows:\\
\begin{enumerate}
    \item Calculate the regular acceleration $\bm{F}_R$ of every particle.
    \item \label{step2} Apply the velocity kick from equation~\ref{eq:kick} to each particle.
    \item Integrate $\bm{F}_I$ using the Hermite integrator to time $t = t + \Delta t_{\text{reg}}$.
    \item Calculate $\bm{F}_R$ again.
    \item Apply another velocity kick.
    \item Go to step~\ref{step2}.
\end{enumerate}%
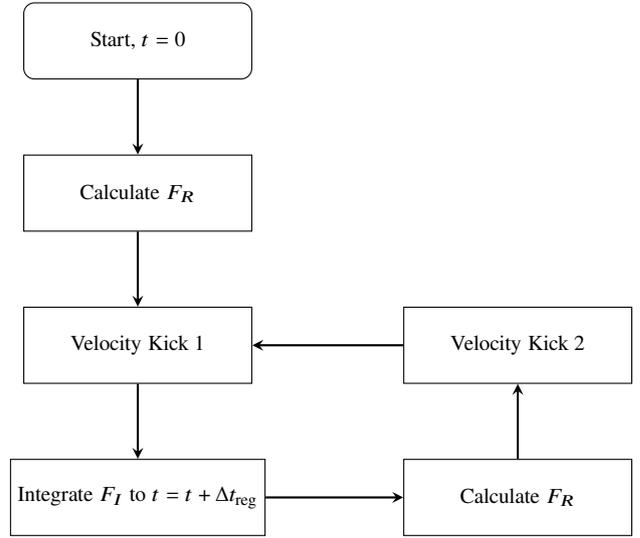
\begin{figure}
    \centering
    \begin{tikzpicture}[node distance=2cm]
\node (start) [start] {Start, $t = 0$};
\node (regf1) [process, below of=start] {Calculate $F_R$};
\node (kick1) [process, below of=regf1] {Velocity Kick 1};
\node (hermite) [process, below of=kick1] {Integrate $F_I$ to $t = t + \Delta t_{\text{reg}}$};
\node (regf2) [process, right of=hermite, xshift=3cm] {Calculate $F_R$};
\node (kick2) [process, above of=regf2] {Velocity Kick 2};
\draw [arrow] (start) -- (regf1);
\draw [arrow] (regf1) -- (kick1);
\draw [arrow] (kick1) -- (hermite);
\draw [arrow] (hermite) -- (regf2);
\draw [arrow] (regf2) -- (kick2);
\draw [arrow] (kick2) -- (kick1);
\end{tikzpicture}
    \caption{The integration algorithm used by \NBPPPT.}
    \label{fig:intgrt}
\end{figure}%
The routine which predicts the position and velocity of a particle at time $t$ uses the total force in \NBODY~in the form

\begin{align}
    \label{eq:nbody-jpred-x}
    \bm{x}_{\text{pred}, i} &= \bm{x}_i + \Delta t_i \bm{\dot{x}}_i + \frac{\Delta t_i^2 (\bm{F}_{I,i} + \bm{F}_{R,i})}{2}  + \frac{\Delta t_i^3 (\bm{\dot{F}}_{I,i} + \bm{\dot{F}}_{R,i})}{6},\\
    \label{eq:nbody-jpred-dx}
    \bm{\dot{x}}_{\text{pred}, i} &= \bm{\dot{x}}_i + \Delta t_i (\bm{F}_{I,i} + \bm{F}_{R,i}) + \frac{\Delta t_i^2 (\bm{\dot{F}}_{I,i} + \bm{\dot{F}}_{R,i})}{2},\\
    \Delta t_i &= t - t_i,
\end{align}%
however due to the velocity kick from the leapfrog routine in \NBPPPT, the regular force would be counted twice. The new routine eschews the regular force and it's derivative, becoming%
\begin{align}
    \label{eq:p3t-jpred-x}
    \bm{x}_{\text{pred}, i} &= \bm{x}_i + \Delta t_i \bm{\dot{x}}_i + \frac{\Delta t_i^2}{2} \bm{F}_{I,i} + \frac{\Delta t_i^3}{6} \bm{\dot{F}}_{I,i},\\
    \label{eq:p3t-jpred-dx}
    \bm{\dot{x}}_{\text{pred}, i} &= \bm{\dot{x}}_i + \Delta t_i \bm{F}_{I,i} + \frac{\Delta t_i^2}{2} \bm{\dot{F}}_{I,i}.
\end{align}%
\subsection{Neighbour Radius}

\NBPPPT~uses a constant neighbour radius $R_s$ for all particles and introduces a new constant parameter, $R_{\text{buff}}$ such that $R_{\text{cut}} = R_s + R_{\text{buff}}$. The fixed neighbour radius leads to a much smaller average neighbour number, usually $< 1$, with most particles being isolated. In contrast to \NBODY~which adjusts size of the neighbour sphere to avoid isolated particles, routines in \NBPPPT~must tolerate empty neighbour lists.

\subsection{Force Calculation}

Historically, the dominant component with regards to computational time in N-body gravitational simulations has been the calculation of the $N^2$ gravitational interactions~\citep{iwasawa2015gpuenabled,Bonsai,aarseth_2003,MIKKOLA1998309,1986Natur.324..446B}. A Barnes-Hut tree~\citep{1986Natur.324..446B} is a data structure and algorithm for approximating the force acting on each particle in $O(n \log n)$ time; the increase in efficiency is a trade-off with accuracy. This trade-off is managed through the opening angle parameter $\theta$. A value of $\theta = 0$ will result in no approximations, reducing the tree code algorithm to the equivalent of a direct-N code. For all runs in this paper we chose $\theta = 0.4$.

The regular force calculator in \NBPPPT~is a GPU-enabled tree code provided by the \program{Bonsai}~library~\citep{Bonsai}. Modifications were made to \program{Bonsai}~to populate the neighbour list and return the results in a format compatible with \NBODY.

Irregular forces, the gravitational force acting on each particle due to its neighbours, are calculated with a family of routines which were modified in \NBPPPT~to include the smoothing function from \eqref{eq:duncan} and \eqref{eq:duncan2} and the alternative prediction formulation from \eqref{eq:p3t-jpred-x} and \eqref{eq:p3t-jpred-dx}.

\subsection{Regularisation Parameters}
\label{reg.params}
\NBODY~defines two values, $\Delta T_{\text{min}}$ and $R_{\text{min}}$, which control when two particles are candidates for KS regularisation. These parameters describe the minimum distance and irregular time step required for two particles to form a regularised pair. Put simply, if for any single particle $i$, $\Delta t_i < \Delta T_{\text{min}}$ and for its nearest neighbour $j$, $|x_i - x_j| < R_{\text{min}}$, then the pair $i,j$ may be removed from the simulation and replaced with a single centre-of-mass particle to represent the binary. Choosing the wrong values for these parameters leads to large energy errors. The values of $R_{\text{min}}$ and $\Delta T_{\text{min}}$ are calculated in the \routine{ADJUST}~routine and are defined as follows.%
\begin{align}
    R_{\text{min}} &= \frac{4 r_h}{N \rho_{\text{core}}^{1/3}},\\
    \Delta T_{\text{min}} &= 0.01 \sqrt{\eta / 0.02} \sqrt{R_{\text{min,}}^3}
\end{align}%
where $r_h$ is the half-mass radius and $\rho_{\text{core}}$ is the central density of the system. 
In \NBPPPT, we use the same definitions for $\Delta T_{\text{min}}$ and $R_{\text{min}}$ as in \NBODY. 

The core quantities, $R_{\text{core}}$, $N_{\text{core}}$, and $\rho_{\text{core}}$ posed a challenge. The method, due to~\citet{1985ApJ...298...80C}, used to find $R_{\text{core}}$ and the core density $\rho_{\text{core}}$ relies on knowing the $6$ nearest neighbours to each particle. \NBODY~uses the neighbour lists of the particles for this purpose and while \NBODY~ensures that most neighbour lists have sufficient numbers by varying $r_{\text{cut}}$ for individual particles, \NBPPPT~uses a global fixed $r_{\text{cut}}$. 

The core quantities are used to determine the regularisation parameters, $R_{\text{min}}$ and $\Delta T_{\text{min}}$. These parameters describe the minimum distance and irregular time step required for two particles to form a regularised pair. If these parameters are too high, \NBPPPT~chooses to regularise pairs in sub-optimal situations, leading to frequent and unnecessary regularisations. Conversely, when the parameters are too low, particles will get too close to each other before they are regularised, leading to small irregular integration steps. We show a solution to this problem in subsection~\ref{collapse.results}.

\subsection{Other Modifications}
Assumptions are present throughout \NBODY, particularly with regards to the variability of $R_s$ and the avoidance of isolated particles. Also prevalent in the code are numerous code loops to perform local Hermite scheme predictions and force calculations, outside of the usual routines, which needed to be modified to follow the formulations from \eqref{eq:p3t-jpred-x} and \eqref{eq:p3t-jpred-dx}. To explain all of the changes required in \NBPPPT~would be prohibitively verbose. Tables listing the modifications made to \NBODY~are made available in appendix~\ref{appendix.tables}.

\section{Results}
\label{results}
\subsection{Accuracy and Performance}
\label{acc.perf}
We performed a number of test runs in order to compare the accuracy and performance of \NBPPPT~against \NBODY. In this section, we describe the configuration of each program and the results of the test runs. For initial conditions, we adopted a Plummer model~\citep{Plummer} of equal-mass particles for each run. We used N-Body units~\citep{1986LNP...267...13H} and set the total mass $M = 1$, the gravitational constant $G = 1$, and the total energy $E = -1/4$. Softening is not used as KS regularisation will avoid the singularity of the gravitational potential.

All runs were performed on a single compute node, utilising $28$ Intel Xeon Gold 6132 CPUs ($2$ sockets, $14$ cores per socket), $32\text{GB}$ of memory, and $1$ NVIDIA V100 GPU. For each run, we evolved the system over $10$ N-body time units, with the exception of $N=2048k$ which was evolved for only $1$ N-body time unit to avoid lengthy run times. In our results we show the average wall clock time per N-body time unit.

\subsubsection{Parameters}

The number of particles, $N$, in each run ranged from $N=32k$ to $2048k$ (where $k = 1024$), doubling in size for each successive run. Unless explicitly stated, run time parameters were chosen to follow the standard \NBODY~scheme. The parameters $r_{\text{buff}}$, $r_{\text{cut}}$, $\Delta t_{\text{reg}}$ were determined according to a modified version of the optimal set of accuracy parameters set out in~\citet{iwasawa2015gpuenabled} and summarised below.%
\begin{align}
    r_{\text{buff}} &= 3 \tau \sigma,\label{eq:rbuff}\\
    r_{\text{cut}} &= 4 \tau,\\
    \tau &= \frac{1}{128} \left(\frac{N}{2^{14}}\right)^{-1/3},\\
    \Delta t_{\text{reg}} &= \text{lowest power of } \frac{1}{2} \mid \Delta t_{\text{reg}} \leq \tau,\\
    \Delta t_{\text{max}} &= \frac{\Delta t_{\text{reg}}}{4},\label{eq:dtmax}\\
    \sigma &= \frac{1}{\sqrt{2}}.
\end{align}%
Here, $\sigma$ is the global three-dimensional velocity dispersion and $\tau$ is the calculated regular time step used to get $r_{\text{buff}}$ and $r_{\text{cut}}$. $\Delta t_{\text{max}}$ identifies the maximum irregular time step for an individual particle in \NBPPPT; for \NBODY~the standard value of $1/8$ was used. 

For all runs in this paper, $\sigma$ is considered constant throughout the run. In future runs, requiring longer integration times, a periodic recalculation $\sigma$ and subsequently of equations \eqref{eq:rbuff} to \eqref{eq:dtmax} would be necessary. 

\subsubsection{Results}

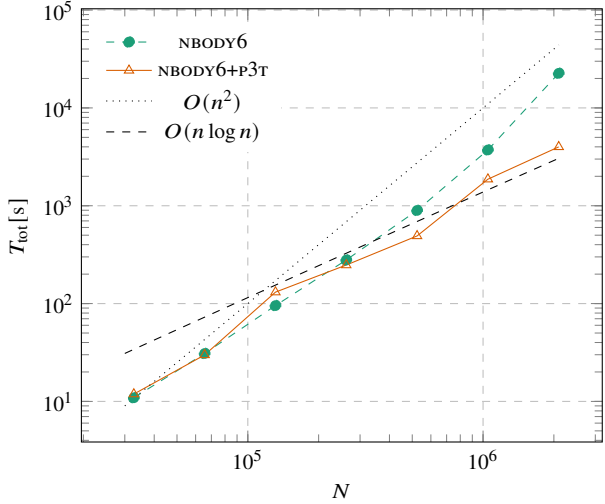
\begin{figure}
    \centering
    \begin{tikzpicture}
\begin{axis}[    
    domain=30000:2100000,
    xlabel={$N$},
    ylabel={$T_{\text{tot}}[\text{s}]$},
    ymode=log,
    xmode=log,
    legend pos=north west,
    legend style ={ draw=none },
    ymajorgrids=true,
    xmajorgrids=true,
    grid style=dashed,
    cycle list name=colour list params
]

\addplot 
    table [x=n, y=tot, col sep=comma] {instrument-gpu.csv};
    
\addplot 
    table [x=n, y=tot, col sep=comma] {instrument-tree.csv};
    
\addplot[dotted, color=black]  {1.0e-8 * x*x};
\addplot[dashed, color=black]  {1e-4 * x*ln(x)};

    \legend{\NBODY,\NBPPPT,$O(n^{2})$,$O(n \log{n})$}
\end{axis}
\end{tikzpicture}
    \caption{\textbf{Total wall-clock time of execution per N-body time unit as a function of $N$.} \NBODY~runs used the same initial conditions as the corresponding \NBPPPT~runs.}
    \label{fig:tot-time}
\end{figure}

When $N \leq 256k$, \NBPPPT~performs similarly to~\NBODY~in terms of run time. For larger runs, \NBPPPT~is measurably faster, completing the run in less than half the time for $N = 1024k$ and more than $5$ times faster for $N = 2048k$. Fig.~\ref{fig:tot-time} shows the evolution of total execution time as the system size increases for both codes. 

We identified four functional components common to both codes: force calculation and integration, each further separated into regular and irregular parts. Irregular and regular force calculation execution time are denoted $T_{\bm{F}_I}$ and $T_{\bm{F}_R}$ respectively. Similarly, irregular and regular integration are denoted $T_I{^\text{int}}$ and $T_R{^\text{int}}$ respectively. Here, the irregular force calculation and integration corresponds to the direct-N force calculator and Hermite integrator of both \NBODY~and \NBPPPT. Furthermore, the regular force calculation corresponds to the GPU-enabled direct-N force calculator in \NBODY~and to the GPU-enabled tree code in ~\NBPPPT. Finally, the regular integration component corresponds to the Hermite integrator in \NBODY~and to the leap frog integrator in \NBPPPT. We measured the execution time of each component to show how the choice of component algorithm affects the total execution time.

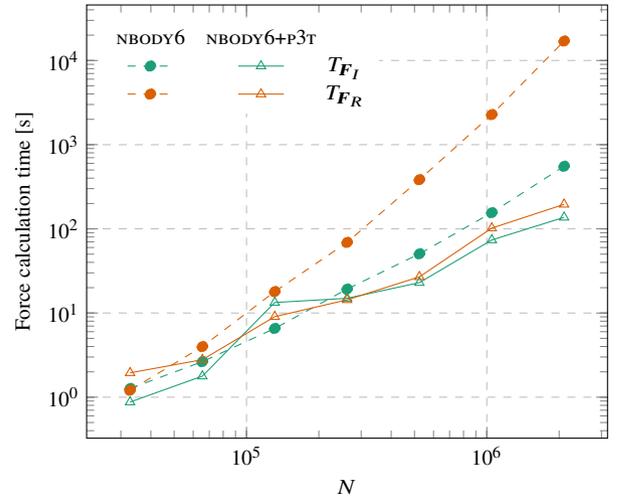
\begin{figure}
    \centering
    \begin{tikzpicture}%

\begin{axis}[
    group style={
        group name=force time plots,
        group size=1 by 2,
        xlabels at=edge bottom,
        xticklabels at=edge bottom,
        vertical sep=0pt
    },
    xlabel={$N$},
    ylabel={Force calculation time [s]},
    ymode=log,
    xmode=log,
    legend columns=2,
    legend pos=north west,
    legend style ={ draw=none },
    ymajorgrids=true,
    xmajorgrids=true,
    grid style=dashed,
    cycle list name=colour list times
]

\addlegendimage{legend image with text=\NBODY}
\addlegendentry{}
\addlegendimage{legend image with text=\NBPPPT}
\addlegendentry{}

\addplot 
    table [x=n, y=f_irr, col sep=comma] {instrument-gpu.csv};
    
\addlegendentry{}

\addplot 
    table [x=n, y=f_irr, col sep=comma] {instrument-tree.csv};

\addlegendentry{$T_{\bm{F}_{I}}$}

\addplot 
    table [x=n, y=f_reg, col sep=comma] {instrument-gpu.csv};
    
\addlegendentry{}

\addplot 
    table [x=n, y=f_reg, col sep=comma] {instrument-tree.csv};
    
\addlegendentry{$T_{\bm{F}_{R}}$}

\end{axis}
\end{tikzpicture}
    \caption{\textbf{Wall-clock time of force calculation as a function of $N$.} Both regular and irregular force calculation times are shown. All runs are the same as those in figure \ref{fig:tot-time}.}
    \label{fig:force-time}
\end{figure}

Fig.~\ref{fig:force-time}~shows  $T_{\bm{F}_I}$ and $T_{\bm{F}_R}$ for different system sizes. \NBPPPT~consistently spends less time calculating either force when $N > 128k$. For smaller $N$ the difference is negligible; only a few seconds per N-body time.

The advantage of the tree code can be seen in Fig.~\ref{fig:force-time}. As $N$ increases the execution time clearly increases much slower in \NBPPPT~than the direct-N algorithm of \NBODY. At $N = 1024k$, the tree code is faster than the direct-N code by approximately a factor of $20$.

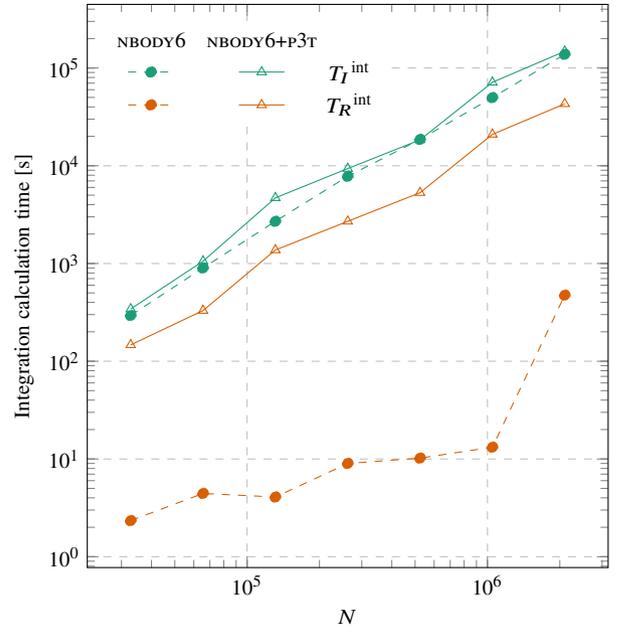
\begin{figure}
    \centering
    \begin{tikzpicture}%

\begin{axis}[
    group style={
        group name=intgrt time plots,
        group size=1 by 2,
        xlabels at=edge bottom,
        xticklabels at=edge bottom,
        vertical sep=0pt
    },
    xlabel={$N$},
    ylabel={Integration calculation time [s]},
    ymode=log,
    xmode=log,
    legend columns=2,
    legend style = { draw = none },
    legend pos = north west,
    y post scale = 1.3,
    ymajorgrids=true,
    xmajorgrids=true,
    grid style=dashed,
    cycle list name=colour list times
]

\addlegendimage{legend image with text=\NBODY}
\addlegendentry{}
\addlegendimage{legend image with text=\NBPPPT}
\addlegendentry{}

\addplot 
    table [x=n, y expr=\thisrowno{2}*60, col sep=comma] {instrument-gpu.csv};
    
\addlegendentry{}
    
\addplot 
    table [x=n, y expr=\thisrowno{2}*60, col sep=comma] {instrument-tree.csv};
    
\addlegendentry{$T_{I}{^\text{int}}$}

\addplot 
    table [x=n, y expr=\thisrowno{3}*60, col sep=comma] {instrument-gpu.csv};
    
\addlegendentry{}

\addplot 
    table [x=n, y expr=\thisrowno{3}*60, col sep=comma] {instrument-tree.csv};
    
\addlegendentry{$T_{R}{^\text{int}}$}

\end{axis}
\end{tikzpicture}
    \caption{\textbf{Wall-clock time of integration as a function of $N$.} Both regular and irregular integration times are shown and do not include force calculation time. All runs are the same as those in figure \ref{fig:tot-time}.}
    \label{fig:intgrt-time}
\end{figure}

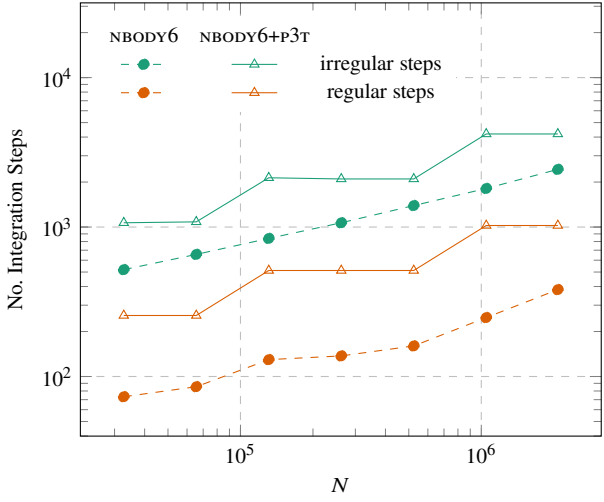
\begin{figure}
    \centering
    \begin{tikzpicture}
\begin{axis}[    
    xlabel={$N$},
    ylabel={No. Integration Steps},
    xmode=log,
    ymode=log,
    legend pos=north west,
    ymajorgrids=true,
    xmajorgrids=true,
    grid style=dashed,
    legend columns=2,
    legend style = { draw = none },
    legend pos = north west,
    ymax = 10^4.5,
    cycle list name=colour list times
]
\addlegendimage{legend image with text=\NBODY}
\addlegendentry{}
\addlegendimage{legend image with text=\NBPPPT}
\addlegendentry{}

\addplot 
    table [x=n, y=gpu-pc, col sep=comma] {nsteps.csv};
\addlegendentry{}

\addplot 
    table [x=n, y=tree-pc, col sep=comma] {nsteps.csv};
\addlegendentry{irregular steps}
    
\addplot 
    table [x=n, y=gpu-pc, col sep=comma] {nsteps-reg.csv};
\addlegendentry{}

\addplot 
    table [x=n, y=tree-pc, col sep=comma] {nsteps-reg.csv};
\addlegendentry{regular steps}

\end{axis}
\end{tikzpicture}
    \caption{\textbf{The average number of integration steps per particle per N-body unit time as a function of $N$.} Both regular and irregular integration steps are shown. All runs are the same as those in figure \ref{fig:tot-time}.}
    \label{fig:nsteps}
\end{figure}

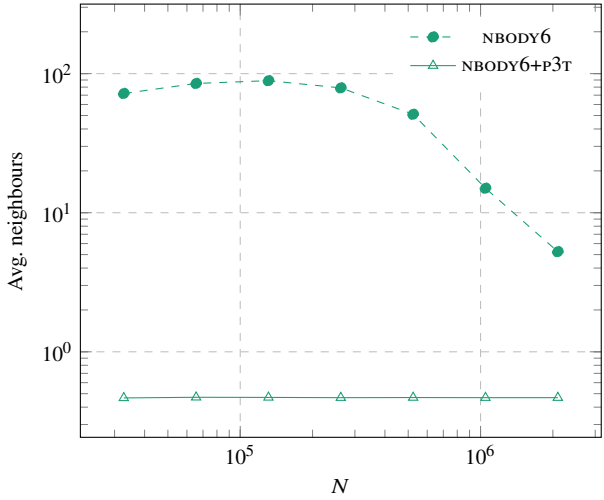
\begin{figure}
    \centering
    \begin{tikzpicture}
\begin{axis}[    
    xlabel={$N$},
    ylabel={Avg. neighbours},
    ymode=log,
    xmode=log,
    ymax={10^2.5},
    legend pos=north west,
    ymajorgrids=true,
    xmajorgrids=true,
    grid style=dashed,
    legend pos=north east,
    legend style ={ draw=none },
    cycle list name=colour list times
]

\addplot 
    table [x=n, y=gpu, col sep=comma] {nnb.csv};

\addplot 
    table [x=n, y=tree, col sep=comma] {nnb.csv};
    
    \legend{\NBODY,\NBPPPT}
\end{axis}
\end{tikzpicture}
    \caption{\textbf{The average number of neighbours per particle at start time as a function of $N$.} All runs are the same as those in figure \ref{fig:tot-time}.}
    \label{fig:nnb}
\end{figure}

In terms of $T_I{^\text{int}}$, \NBPPPT~performs comparably with \NBODY~as seen in Fig.~\ref{fig:intgrt-time}. Fig.~\ref{fig:nsteps} shows that \NBPPPT~performs slightly more irregular integration steps than \NBODY, but Fig.~\ref{fig:nnb} shows that particles in \NBPPPT~have far fewer neighbours on average which would lead to less time spent calculating irregular forces per particle. Both programs also use the same irregular integration algorithm.

Fig.~\ref{fig:intgrt-time} also shows that \NBPPPT~spends more time performing regular integration $T_R{^\text{int}}$ than \NBODY. The regular integration algorithms used by each program are not comparable, making it difficult to diagnose the disparity. The total number of regular steps performed by \NBPPPT~is fixed at $N / \Delta t_{\text{reg}}$ per unit time. The number of regular steps performed by \NBODY~is dependent upon the Hermite algorithm which drives regular integration. Fig.~\ref{fig:nsteps} shows that \NBPPPT~performs more regular steps per particle than \NBODY, over four times as many for $N > 10^6$, which would explain why \NBPPPT~spends more time overall performing regular integration than \NBODY. 

The average number of neighbours in \NBPPPT~is much smaller, which is reflected in the irregular force calculation time. However, the total integration time is slower than in \NBODY~because the number of irregular and regular steps per particle is higher due to the smaller maximum time step value. In larger system sizes this is not so obvious because the regular force calculation is much more computationally expensive.

\begin{figure}
    \centering
    \begin{tikzpicture}%

\begin{groupplot}[
    group style={
        group name=all time plots,
        group size=1 by 2, 
        xlabels at=edge bottom,
        xticklabels at=edge bottom,
        vertical sep=0pt
    },
    domain=30000:2100000,
    xlabel={$N$},
    ylabel={$T / T_{\text{tot}}$},
    ymode=log,
    xmode=log,
    xmin=3E+4,
    xmax=2.3E+6,
    ymin=10^-5.5,
    legend style ={ draw=none },
    legend pos = south west,
    legend columns = 3,
    ymajorgrids=true,
    xmajorgrids=true,
    grid style=dashed,
    cycle list name=colour list times all,
]
\nextgroupplot
\addplot
    table [x=n, y=intgrt_irr, col sep=comma] {instrument-norm-gpu.csv};
\addplot
    table [x=n, y=intgrt_reg, col sep=comma] {instrument-norm-gpu.csv};
\addplot
    table [x=n, y=f_irr, col sep=comma] {instrument-norm-gpu.csv};
\addplot
    table [x=n, y=f_reg, col sep=comma] {instrument-norm-gpu.csv};
\addplot
    table [x=n, y expr=1.0 - (\thisrowno{1} + \thisrowno{2} + \thisrowno{3} + \thisrowno{4}), col sep=comma] {instrument-norm-gpu.csv};
    
    \legend{
        $T{^\text{int}}_{I} / T_{\text{tot}}$,
        $T{^\text{int}}_{R} / T_{\text{tot}}$,
        $T_{F_{I}} / T_{\text{tot}}$,
        $T_{F_{R}} / T_{\text{tot}}$,
        $T_{\text{List}}$
    }

\nextgroupplot[ymin=10^-1.8]

\addplot
    table [x=n, y=intgrt_irr, col sep=comma] {instrument-norm-tree.csv};
\addplot
    table [x=n, y=intgrt_reg, col sep=comma] {instrument-norm-tree.csv};
\addplot
    table [x=n, y=f_irr, col sep=comma] {instrument-norm-tree.csv};
\addplot
    table [x=n, y=f_reg, col sep=comma] {instrument-norm-tree.csv};
\addplot
    table [x=n, y expr=1.0 - (\thisrowno{1} + \thisrowno{2} + \thisrowno{3} + \thisrowno{4}), col sep=comma] {instrument-norm-tree.csv};
    
    \legend{
        $T{^\text{int}}_{I} / T_{\text{tot}}$,
        $T{^\text{int}}_{R} / T_{\text{tot}}$,
        $T_{F_{I}} / T_{\text{tot}}$,
        $T_{F_{R}} / T_{\text{tot}}$,
        $T_{\text{List}}$
    }
\end{groupplot}
\end{tikzpicture}
    \caption{\textbf{Proportion of total wall-clock time spent performing each task, as a function of $N$.} The upper panel is for \NBODY~runs and the lower panel is for \NBPPPT~runs. Included is the value $T_{\text{List}}$ which represents particle scheduling. All runs are the same as those in figure \ref{fig:tot-time}.}
    \label{fig:all-time-proportion}
\end{figure}
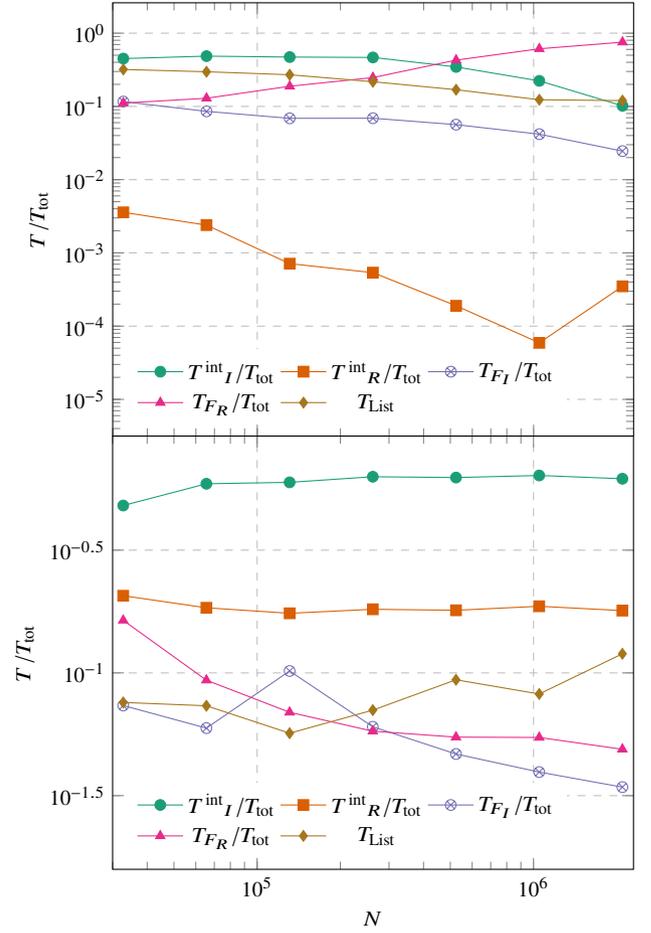

Fig.~\ref{fig:all-time-proportion} shows the execution time of each component as a proportion of the total execution time for that run. Included in this figure is the time $T_{\text{List}}$, which measures the time needed for the particle scheduler to determine which particles are due to be integrated at each step. As $N$ gets large in \NBODY, the regular force calculation, $T_{\bm{F}_R}$, becomes the dominant component accounting for nearly all of the computation time.

However, in \NBPPPT, $T_{\bm{F}_R}$ is no longer dominant. In fact, irregular integration $T_I{^\text{int}}$ becomes the dominant component accounting for approximately $50\%$ to $60\%$ of the run time consistently. The total time spent performing both irregular and regular integration is consistently $70\%$ to $80\%$ of the total run time. Moreover, the proportion of time spent performing force calculations, both regular and irregular, decreases significantly as $N$ gets large; less that $10\%$ altogether at $N=2048k$. 

Both \NBODY~and \NBPPPT~spend around $10\%$ of the run time in $T_{\text{List}}$, however this proportion is decreasing in \NBODY~and increasing in \NBPPPT. Both codes use the same algorithm; first the particles with small time steps are copied in to a smaller list $L_Q$ at regular intervals. Next, the smaller list is scanned at every block step to determine which particles are due at that step. The first part has a run time complexity of $O(N)$, and the second is $O(N_Q)$ where $N_Q$ is the size of $L_Q$. If $L_Q$ is repopulated too frequently, or if $N_Q$ is too often larger than the actual number of particles due to be integrated, then it is possible for $T_\text{List}$ to take longer than $T_I{^\text{int}}$. The distributions of time steps are different in \NBODY~and \NBPPPT, which leads to different measurements of $T_\text{List}$ in each. In \NBPPPT~it is possible to replace the scheduling technique described above with a priority queue, however it is beyond the scope of this paper.

\subsection{Core Collapse}
As a proof-of-concept, we performed a long run using \NBPPPT~to the initial core collapse. The cluster, containing $64k$ particles, was arranged as a Plummer sphere with the following mass spectrum: %
\begin{equation}
    m_{i} =  \begin{cases}
               \frac{10}{1.9 N} &\text{if }i \equiv 0 \pmod{10},\\
               \frac{1}{1.9 N} &\text{otherwise}.
            \end{cases}
\end{equation}%
In this way, approximately half of the system's mass is contained in just $10\%$ of the particles. All other parameters were the same as in section \ref{acc.perf}. 

An \NBODY~run with identical initial conditions was performed to compare the evolution of system properties. Lagrange radii for a range of mass proportions, total energy error ($\frac{|E(t) - E_0|}{E_0}$), core density ($\rho_{\text{core}}$), core radius ($R_{\text{core}}$), core membership ($N_{\text{core}}$), and the KS regularisation parameters $R_{\text{min}}$ and $\Delta T_{\text{min}}$ were all measured in both runs.

\subsubsection{Results}
\label{collapse.results}

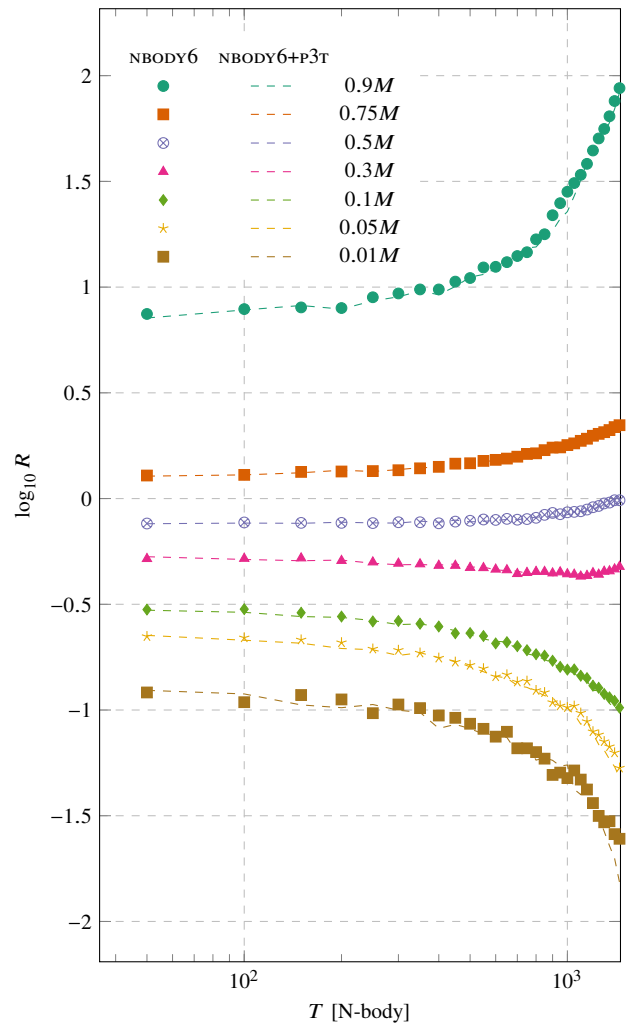
\begin{figure}
    \centering
    \begin{tikzpicture}%
\begin{axis}[
    xmode=log,
    xmax=1460,
    xlabel={$T$ [N-body]},
    ylabel={$\log_{10} R$},
    ymajorgrids=true,
    xmajorgrids=true,
    grid style=dashed,
    cycle list name=colour list corr,
    legend columns=2,
    legend style = { draw = none },
    legend pos = north west,
    y post scale = 2.2
]

\addlegendimage{legend image with text=\NBODY}
\addlegendentry{}
\addlegendimage{legend image with text=\NBPPPT}
\addlegendentry{}

\addplot 
    table [x=T, y=0.9, col sep=comma, each nth point=5] {lagr-gpu.csv};
\addlegendentry{}
\addplot 
    table [x=T, y=0.9, col sep=comma, each nth point=5] {lagr-tree.csv};
\addlegendentry{$0.9 M$}

\addplot 
    table [x=T, y=0.75, col sep=comma, each nth point=5] {lagr-gpu.csv};
\addlegendentry{}
\addplot 
    table [x=T, y=0.75, col sep=comma, each nth point=5] {lagr-tree.csv};
\addlegendentry{$0.75 M$}
    
\addplot 
    table [x=T, y=0.5, col sep=comma, each nth point=5] {lagr-gpu.csv};
\addlegendentry{}
\addplot 
    table [x=T, y=0.5, col sep=comma, each nth point=5] {lagr-tree.csv};
\addlegendentry{$0.5 M$}
    
\addplot 
    table [x=T, y=0.3, col sep=comma, each nth point=5] {lagr-gpu.csv};
\addlegendentry{}
\addplot 
    table [x=T, y=0.3, col sep=comma, each nth point=5] {lagr-tree.csv};
\addlegendentry{$0.3 M$}

\addplot 
    table [x=T, y=0.1, col sep=comma, each nth point=5] {lagr-gpu.csv};
\addlegendentry{}
\addplot 
    table [x=T, y=0.1, col sep=comma, each nth point=5] {lagr-tree.csv};
\addlegendentry{$0.1 M$}

\addplot 
    table [x=T, y=0.05, col sep=comma, each nth point=5] {lagr-gpu.csv};
\addlegendentry{}
\addplot 
    table [x=T, y=0.05, col sep=comma, each nth point=5] {lagr-tree.csv};
\addlegendentry{$0.05 M$}

\addplot 
    table [x=T, y=0.01, col sep=comma, each nth point=5] {lagr-gpu.csv};
\addlegendentry{}
\addplot 
    table [x=T, y=0.01, col sep=comma, each nth point=5] {lagr-tree.csv};
\addlegendentry{$0.01 M$}

\end{axis}
\end{tikzpicture}
    \caption{Lagrange radii at different proportions of the total mass $M$. The solid and dashed lines represent the \NBODY~and \NBPPPT~runs, respectively. Both runs used the the same initial conditions; $N=64k$ particles in a Plummer sphere with a mass spectrum.}
    \label{fig:lagrange-radii}
\end{figure}

\begin{figure}
    \centering
    \begin{tikzpicture}
\begin{loglogaxis}[    
    xlabel={$T$ [N-body]},
    ylabel={$|E(t) - E_0| / E_0$},
    xmax=1460,
    xmin=1,
    domain=1:1460,
    ymajorgrids=true,
    xmajorgrids=true,
    grid style=dashed,
    legend pos=north west,
    legend style ={ draw=none },
    cycle list name=colour list params,
]

\addplot 
    table [x index=0, y expr=abs(\thisrowno{1}), col sep=comma, each nth point=25] {gpu-ene.csv};
\addplot 
    table [x index=0, y expr=abs(\thisrowno{1}), col sep=comma, each nth point=25] {tree-ene.csv};
    
\addplot[black,no marks] {6.47108e-06 * x + 0.0016388};
\addplot[dashed,no marks] {3.06373e-06 * x + 3.018e-07};
    
    \legend{
        \NBODY,
        \NBPPPT,
        \NBODY~fit,
        \NBPPPT~fit
    }
\end{loglogaxis}
\end{tikzpicture}
    \caption{Energy change as a function of time as a proportion of the initial energy. Both runs are the same as those in figure~\ref{fig:lagrange-radii}.}
    \label{fig:energy}
\end{figure}
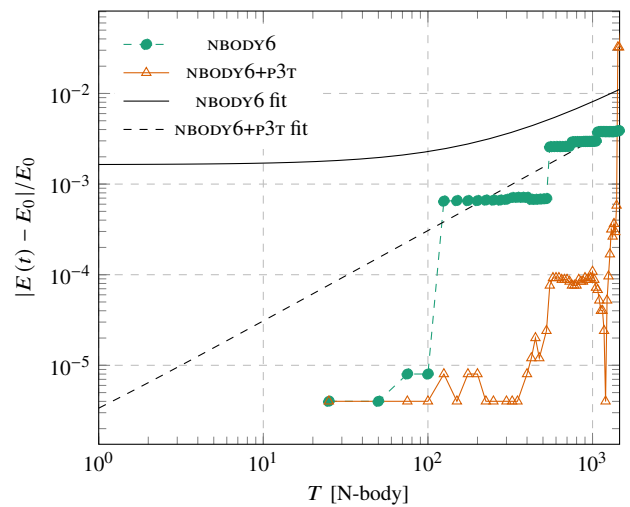

Fig.~\ref{fig:lagrange-radii} shows the evolution of the Lagrange radii to the initial core collapse. Both runs exhibit near-identical overall evolution for the measured mass proportions. Additionally, Fig.~\ref{fig:energy} shows the total energy change as a function of time; the \NBPPPT~run recorded an error rate of the same order of magnitude as that of the \NBODY~run. A linear regression over both series gave a slope of $~6.47 \times 10^{-6} \pm 1.27 \times 10^{-7} E/T$ and $~3.06 \times 10^{-6} \pm 3.02 \times 10^{-7} E/T$ for the \NBODY~and~\NBPPPT~runs, respectively.

As was mentioned in subsection~\ref{reg.params}, a fixed neighbour radius can be problematic if too many particles do not have enough neighbours to contribute to the calculation of the core quantities. Fig.~\ref{fig:nnb} shows that the average number of neighbours per particle in \NBPPPT~is quite low. In fact, few particles have the requisite $6$ neighbours required to contribute to the calculations of $R_{\text{core}}$ and $N_{\text{core}}$.

A new algorithm was introduced to \NBPPPT~to accurately approximate the $R_{\text{core}}$, $N_{\text{core}}$, and $\rho_{\text{core}}$ values. Instead of using the neighbour list of each particle to determine its $6$ nearest neighbours, our code takes a random sample of $\min({N, \sqrt{5000 N}})$ particles, then performs a brute-force search for those particles' $6$ nearest neighbours.

\begin{figure}
    \centering
    \begin{tikzpicture}
\begin{groupplot}[
    group style={
        group name=params plots,
        group size=1 by 2,
        xlabels at=edge bottom,
        xticklabels at=edge bottom,
        vertical sep=14pt
    },
    xlabel={$T$ [N-body]},
    ylabel={$\Delta T_{\text{min}}$},
    xmode=log,
    xmax=1460,
    ymajorgrids=true,
    xmajorgrids=true,
    grid style=dashed,
    legend style ={ draw=none },
    legend pos = south west,
    cycle list name=colour list params,
]

\nextgroupplot[
    ylabel={$R_{\text{min}}$}
]
\addplot 
    table [x index=0, y index=1, col sep=comma, each nth point=10] {gpu-params.csv};
\addplot 
    table [x index=0, y index=1, col sep=comma, each nth point=10] {tree-params.csv};
    
\legend{\NBODY,\NBPPPT}
    
\nextgroupplot[
    ylabel={$\Delta T_{\text{min}}$}
]
\addplot 
    table [x index=0, y index=2, col sep=comma, each nth point=10] {gpu-params.csv};
\addplot 
    table [x index=0, y index=2, col sep=comma, each nth point=10] {tree-params.csv};
    
    \legend{\NBODY,\NBPPPT}
\end{groupplot}
\end{tikzpicture}
    \caption{The KS regularisation parameters, $R_{\text{min}}$ and $\Delta T_{\text{min}}$ shown in the top and bottom panels, respectively. Both runs are the same as those in figure~\ref{fig:lagrange-radii}.}
    \label{fig:dtmin}
\end{figure}
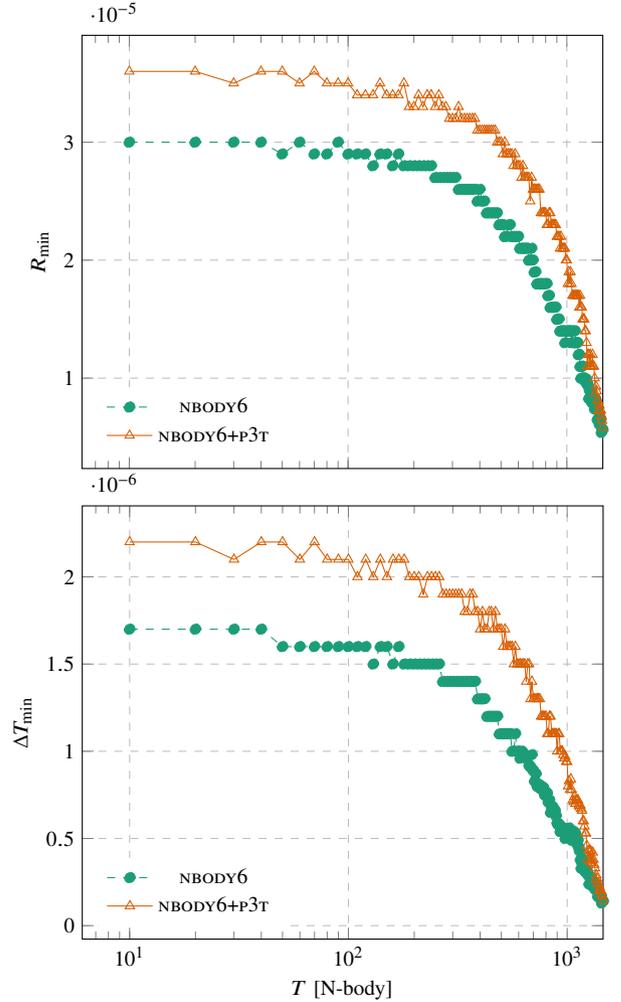

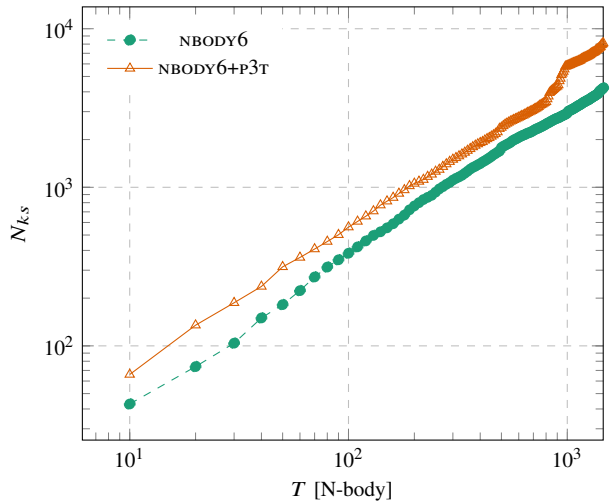
\begin{figure}
    \centering
    \begin{tikzpicture}
\begin{axis}[
    xlabel={$T$ [N-body]},
    ylabel={$N_{ks}$},
    ymode=log,
    xmode=log,
    xmax=1460,
    ymajorgrids=true,
    xmajorgrids=true,
    grid style=dashed,
    legend style ={ draw=none },
    legend pos = north west,
    cycle list name=colour list params
]

\addplot 
    table [x index=0, y index=1, col sep=comma, each nth point=1] {gpu-nks.csv};
\addplot 
    table [x index=0, y index=1, col sep=comma, each nth point=1] {tree-nks.csv};
    
    \legend{\NBODY,\NBPPPT}
\end{axis}
\end{tikzpicture}
    \caption{The number of KS pairs formed as a function of time . Both runs are the same as those in figure~\ref{fig:lagrange-radii}.}
    \label{fig:nks}
\end{figure}

Figures~\ref{fig:core},~\ref{fig:core-n}, and~\ref{fig:core-density} respectively show the changes in these values as a function of time using this new algorithm. With the core quantities correctly determined, then regularisation parameters can be confidently calculated. Fig.~\ref{fig:dtmin} shows both $R_{\text{min}}$ and $\Delta T_{\text{min}}$ calculations; \NBPPPT~maintains small offsets from the output of \NBODY, but they converge to the same solutions during core collapse when correct regularisation parameters are critical. Furthermore, Fig.~\ref{fig:nks} shows the number of KS pairs formed as a function of time for each program. \NBPPPT~produces roughly four times as many pairs, and slightly more again during the later stages of the run; this may indicate that the new core quantity algorithm requires some adjustment.

\begin{figure}
    \centering
    \begin{tikzpicture}
\begin{axis}[
    xlabel={$T$ [N-body]},
    ylabel={$R_\text{core}$},
    ymode=log,
    xmode=log,
    xmax=1460,
    ymajorgrids=true,
    xmajorgrids=true,
    grid style=dashed,
    legend style ={ draw=none },
    legend pos = south west,
    cycle list name=colour list params,
]

\addplot 
    table [x index=0, y index=1, col sep=comma, each nth point=5] {gpu-core.csv};
\addplot 
    table [x index=0, y index=1, col sep=comma, each nth point=5] {tree-core.csv};
    
    \legend{\NBODY,\NBPPPT}
\end{axis}
\end{tikzpicture}
    \caption{The core radius of the cluster as a function of time. Both runs are the same as those in figure~\ref{fig:lagrange-radii}.}
    \label{fig:core}
\end{figure}
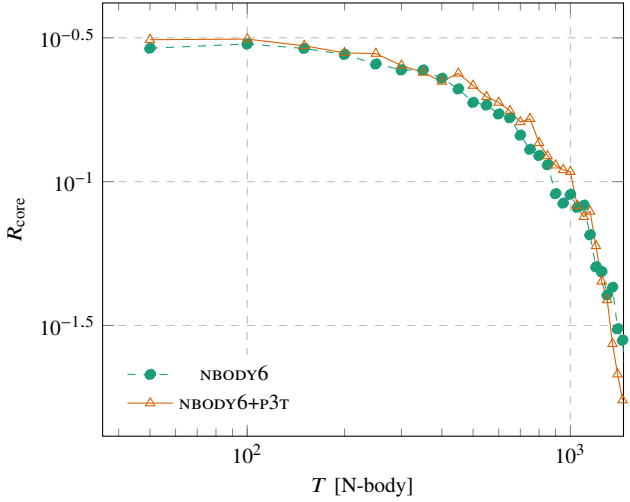
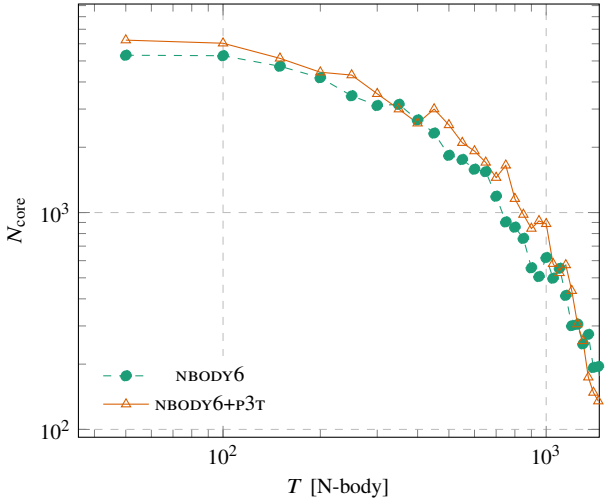
\begin{figure}
    \centering
    \begin{tikzpicture}
\begin{axis}[
    xlabel={$T$ [N-body]},
    ylabel={$N_\text{core}$},
    ymode=log,
    xmode=log,
    xmax=1460,
    ymajorgrids=true,
    xmajorgrids=true,
    grid style=dashed,
    legend style ={ draw=none },
    legend pos = south west,
    cycle list name=colour list params
]

\addplot 
    table [x index=0, y index=2, col sep=comma, each nth point=5] {gpu-core.csv};
\addplot 
    table [x index=0, y index=2, col sep=comma, each nth point=5] {tree-core.csv};
    
    \legend{\NBODY,\NBPPPT}
\end{axis}
\end{tikzpicture}
    \caption{The number of particles in the core of the cluster as a function of time. Both runs are the same as those in figure~\ref{fig:lagrange-radii}.}
    \label{fig:core-n}
\end{figure}
\begin{figure}
    \centering
    \begin{tikzpicture}
\begin{axis}[
    xlabel={$T$ [N-body]},
    ylabel={$\rho_\text{core}$},
    ymode=log,
    xmode=log,
    xmax=1460,
    ymajorgrids=true,
    xmajorgrids=true,
    grid style=dashed,
    legend style ={ draw=none },
    legend pos = north west,
    cycle list name=colour list params,
]

\addplot 
    table [x index=0, y index=1, col sep=comma, each nth point=5] {gpu-rhod.csv};
\addplot 
    table [x index=0, y index=1, col sep=comma, each nth point=5] {tree-rhod.csv};
    
    \legend{\NBODY,\NBPPPT}
\end{axis}
\end{tikzpicture}
    \caption{The core density of the cluster as a function of time. Both runs are the same as those in figure~\ref{fig:lagrange-radii}.}
    \label{fig:core-density}
\end{figure}

\section{Conclusions}
\label{conclusions}

We described a modified \NBODY~version to incorporate a GPU-enabled P3T scheme. Using the optimal set of accuracy parameters outlined in~\citet{iwasawa2015gpuenabled}, our code outperformed \NBODY~for system sizes $N \ge 512k$.

For longer runs, we found that the \NBPPPT~code is sufficiently accurate by comparing the total energy conservation and Lagrange radii to \NBODY~for the same initial conditions up to core collapse. At present,~\NBPPPT~is not capable of post-collapse simulations due to large errors in the energy conservation at this point. Future work may include the adjustment of parameters over time to mitigate post-collapse energy errors.

Our faster code provides the ability to accurately simulate the evolution of dense star clusters, like globular clusters or galactic nuclei, over long time scales (one Hubble time.) A realistic cluster could contain $10^6$ stars, making it advantageous to use our \NBPPPT~code. Simulations of these systems can be used to study the conditions of black hole and neutron star mergers~\citep{2016MNRAS.458.1450W}.

We found that the calculation of the KS regularisation parameters needed to be modified due to an inaccurate calculation of the core density $\rho_{\text{core}}$, which in turn was caused by the lower average neighbour number due to a fixed $r_\text{cut}$ value. We have mitigated the issue with a new core density algorithm, although this algorithm may need further adjustment, depending on the type of system that is being simulated.

Additional future improvements may include scaling the GPU-enabled force calculator to utilise multiple GPUs, and to scale the computation over many nodes. A method to deal with higher ratios of binaries, like the one in~\citet{Wang_2020}, would also be useful.

\section*{Acknowledgements}

Parts of this work were performed on the OzSTAR national facility at Swinburne University of Technology. The OzSTAR program receives funding in part from the National Collaborative Research Infrastructure Strategy (NCRIS) Astronomy allocation provided by the Australian Government.

L.W. thanks the financial support from JSPS International Research Fellow (School of Science, The University of Tokyo).

\section*{Data Availability}

The (benchmark) data underlying this article were generated by using the \NBODY~and~\NBPPPT~codes at the supercomputer, Getafix, at the University of Queensland. The data underlying this article will be shared on reasonable request to the corresponding author. 

Our \NBPPPT~code is free for use and is available at \url{https://github.com/anthony-arnold/nbody6-p3t/tree/v1.0.0}.



\bibliographystyle{mnras}
\bibliography{p3t} 




\appendix

\section{Compilation}

\NBODY~traditionally uses the \program{GNU Make} program to build executables. The \NBPPPT~environment is more complex, with the use of \program{Bonsai} as a shared library and the need to ensure that compilation and linkage flags and \program{CUDA} versions remain consistent. To facilitate consistency, \NBPPPT~uses the \program{CMake} program to build libraries and executables.

\section{Code Modifications}
\label{appendix.tables}

 Table~\ref{tab:changes} lists the routines from \NBODY~which were updated in \NBPPPT~along with a short description of what changes were necessary and Table~\ref{tab:new} lists the new routines that were written for \NBPPPT. Table~\ref{tab:vars} describes the new common variables introduced to the program.\\

\begin{table}
\centering
  \caption{Modifications made to \NBODY~routines for \NBPPPT.}
  \begin{tabular}{ll} 
    \hline
    \routine{ADJUST} & Avoid modifying the neighbour radius or regular step.\\
    \hline
    \routine{CORE} & Determine core quantities using a nearest neighbour search of a sample of   random particles.\\
    \hline
    \routine{CHFIRR},\\ \routine{CMFIRR}, \\ \routine{CMFIRR2},\\ \routine{CPERTJ},\\ \routine{CPERTX},\\ \routine{FCHAIN},\\ \routine{FIRR},\\ \routine{FPERT},\\ \routine{KCPERT},\\ \routine{KSPERT},\\ \routine{KSPOLY} & Apply the smoothing functions to calculated forces.\\ 
    \hline
    \routine{FPOLY0} & Read additional parameters and set constant neighbour radius.\\
    \hline
    \routine{INTGRT} & Replace the regular Hermite scheme with a leapfrog scheme.\\
    \hline
    \routine{JPRED},\\ \routine{JPRED2}, \\\routine{KSPRED},\\ \routine{RESOLV},\\ \routine{XCPRED},\\ \routine{XVPRED} & Calculate predictions from equations \eqref{eq:nbody-jpred-x} and \eqref{eq:nbody-jpred-dx}.\\
    \hline
    \routine{KSINIT} & Apply smoothing functions. Avoid modifying the neighbour radius or regular step.\\
    \hline
    \routine{KSTERM} & Use the constant neighbour radius.\\
    \hline
    \routine{MYDUMP} & Read and write new common variables.\\
    \hline
    \routine{NBINT},\\ \routine{NBINTP} & Handle empty neighbour lists and $\bm{F}_{R,i} = 0$.\\
    \hline
    \routine{NBLIST} & Use the constant neighbour radius. Accept empty neighbour lists.\\
    \hline
    \routine{OFFSET} & Offset the regular block time variable \parameter{TBLCKR}.\\
    \hline
    \routine{REGF} & Use \program{Bonsai} to calculate regular forces and determine the neighbour list.\\
    \hline
    \routine{STEPS} & Use a constant regular time step $\Delta t_{\text{reg}}$. Set $\Delta t_i = \Delta t_{\text{max}}$ if $\bm{F}_{R,i} = 0$. \\
    \hline
    \routine{TPERT} & $\Delta t_i = \Delta t_{\text{max}}$ if the neighbour list is empty.\\
    \hline
  \end{tabular}
  \label{tab:changes}
\end{table}

\begin{table}
	\centering
  \caption{New routines in \NBPPPT}
  \begin{tabular}{ll}
    \hline
    \routine{CUTOFF} & Calculate $\bm{F}_{I,i}$ and $\bm{\dot{F}}_{I,i}$ from equations \eqref{eq:fi} and \eqref{eq:fidot}.\\
    \routine{CUTOFF0} & Calculate $\bm{F}_{I,i}$ from equation \eqref{eq:fi}.\\
    \routine{GPUINT} & Apply external forces from \routine{XTRNLF} and \routine{FCLOUD} after \routine{REGF}.\\
    \routine{GPUINT2} & Apply a velocity kick via \routine{RKICK} and determine new time steps.\\
    \routine{RKICK} & Apply a velocity kick from equation \eqref{eq:kick}.\\
    \hline
  \end{tabular}
  \label{tab:new}
\end{table}

\begin{table}
	\centering
  \caption{New common variables in \NBPPPT}
\begin{tabular}{ll}
    \hline
    \parameter{RBUFF} & The parameter $r_{\text{buff}}$.\\
    \parameter{TBLCKR} & Used to track the time of the last regular for calculation.\\
    \parameter{NRMUL} & A constant factor to determine the regular time step; $\Delta t_{\text{reg}} = \texttt{NRMUL} \times \Delta t_{\text{max}}$. Defaults to $4$.\\
    \hline
\end{tabular}
  \label{tab:vars}
\end{table}


\bsp	
\label{lastpage}
\end{document}